# On the Comparability of Chemical Structure and Roughness of Nanochannels in Altering Fluid Slippage

Chinmay Anand Misra and Chirodeep Bakli*

Department of Mechanical Engineering, Indian Institute of Technology Kharagpur, INDIA

*cbakli@gmail.com

**ABSTRACT**

Interfacial hydrodynamic slippage of water depends on both on surface chemistry and roughness. This study tries to connect the effect of chemical property and the physical structure of the surface on the interfacial slippage of water. By performing molecular dynamics simulations (MDS) of Couette flow of water molecules over a reduced Lennard-Jones (LJ) surface, the velocity profile is obtained and extrapolated to get the slip lengths. The slip lengths are measured for various surface-fluid interactions. These interactions are varied by changing the wettability of the surface (characterized by the static contact angle) and its roughness. The slip length variation with the static contact angle as $(1+cos\theta)^{-2}$ is observed. However, it is also observed that the presence of surface roughness always reduces the slip length and it is proposed that the slip length varies with non-dimensionalized average surface roughness as $(1+\alpha^*)^{-2}$. Thus a relation between the chemical wettability and the physical roughness is established and their coupled interactions modifying slip length is probed.

**Keywords:** Wettability, Slip length, Roughness

**Abbreviations:**

NEMS: Nano Electro Mechanical Systems, LJ: Lennard-Jones, MDS: Molecular Dynamics Simulations, fcc: Face Centered Cubic, SPC/E: Simple Point Charge/ Extended

**Introduction**

The flow physics of fluid transport through nanochannels is extremely interesting and intriguing. The surface forces overweigh the volumetric forces at such small length scales, leading to an important role played by the interfacial effects. Thus modifying the surface characteristics of substrates of such channels over nanometer dimensions provides the option to engineer such flows. Study of these flow variations over sub-micron scale is warranted by the advent of research in the areas of Nano Electro Mechanical Systems (NEMS) and Nanoporous Energy Absorption System (NEAS) and also by the urge to study flow of natural fluids in biological membranes. Through extensive molecular dynamics simulations, researchers have explored various aspects of the transport processes and surface interactions concerned with micro and nano scale flows. They have studied the effects of the interface wettability on flow structure, fluid dynamics in surface-nanostructured channels [1], effects of wall lattice-fluid interactions on the density and velocity profiles [2], perturbations in fully developed pressure driven flows [3], slip behavior on substrates with patterned wettability [4], and effects of surface roughness and interface wettability on nanoscale flows [5].

Understanding fluid flow on nano-scale is crucial for designing microfluidic devices, modern developments of nanotechnology like the lab-on-a-chip [6,7], as well as for various applications of porous materials, fluid flow through pores in bio-membranes [8,9], etc. These works have shown that contrary to the macroscopic hydrodynamic theory, the no slip boundary condition might not necessarily hold true for nano-scale channels. In such cases, the fluid velocity $v$ at the surface is more aptly described by a partial slip boundary condition that relates the fluid velocity $v$ at the surface to its gradient $\partial v / \partial z$ in the direction normal to the surface by $v = b(\partial v / \partial z)$, where $b$ is the slip length [10–12]. This interfacial slip has immense practical applications in microfluidics, bio-fluid dynamics, and lubrication etc. by virtue of the fact that it reduces viscous friction at the surfaces and amplifies flow rates in pressure driven flows and electro-osmotic flows [13,14]. This also provides potential to generate power from nano-scale devices [15,16]. Thus it becomes imperative to estimate or measure slip length and study its dependence on interfacial parameters. However, there is a lot of disparity between the experimental data reported for the slip length of typical hydrophobic surfaces. The experimental values of slip lengths have been measured ranging from nanometers [17,18] to micrometers [19]. Researchers have explained larger values of slip length to be due to nano-bubble formation at the interface [20], and molecular dynamics simulations of a model LJ system have shown an increase in slip due to this phenomenon. In an attempt to resolve the controversy in the experimental literature, Huang et al have reported a quasiuniversal dependence of the slip length $b$ on the contact angle $\theta_c$ by performing MDS [21]. Although Huang et al have considered rough surfaces in their study they have not analyzed the dependence of slip length on surface roughness. In this work, it is aimed to establish a mathematical relationship between slip length and average surface roughness by performing non-equilibrium MDS of Couette flow of SPC/E rigid simple point charge model of water between two parallel plates made of reduced LJ atoms. By applying a shear boundary condition, the velocity profile is obtained and extrapolated to get the slip length. In order to mimic the atomically rough surface, the attractive part of LJ interaction potential between fluid and surface [22] is modified. To estimate the contact angle, we perform an equilibrium MDS of a droplet of water on the same surface as used in Couette flow simulation. As a result of this work, it is observed that the slip length varies with non-dimensionalized average surface roughness as $b^* = (1+\alpha^*)^{-2}$. On the basis of this work, a correction in the formula presented by Huang et al [21] to account for variation in surface roughness is proposed.

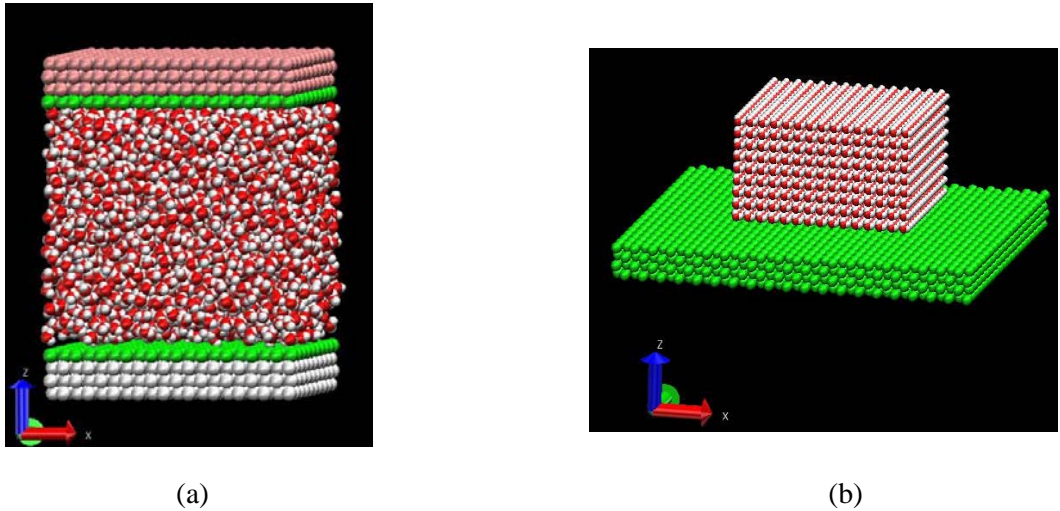

(a) (b)

**Figure 1.** Three dimensional view of the simulation system for (a) Couette Flow (N=2, $\varepsilon_{ww} = 0.4$) and (b) Contact angle estimation (Smooth, $\varepsilon_{ww} = 0.4$)

**Problem Formulation**

The simulation system, for Couette flow simulation, is similar to that used by Huang et al [21]and is shown in figure 1. A channel of height 5nm, bounded by walls made of four layers of reduced LJ atoms arranged in close packed fcc (face-centered cubic) lattice oriented such that (100) face is in contact with water is considered. The lateral dimensions of the walls are $5nm \times 5nm$. Wall atoms interact with each other through a LJ potential. The LJ interaction parameter between wall atoms $\sigma_{ww}$ is taken to be $0.35nm$ while $\epsilon_{ww}$ is varied to change the wall fluid interaction and hence the contact angle. The channel is filled with 3584 water molecules SPC/E model( Simple Point Charge/ Extended). This number ensures that the density of water is $1000 kg/m^3$ for the particular size of the channel considered. Heteronuclear LJ interactions are determined by standard Lorentz-Berthelot combining rules [23]. Roughness is introduced by modifying the attractive part of the LJ potential as shown in Eq.(1).

$$U_{LJ} = 4\epsilon_{wl}\left[\left(\frac{\sigma_{wl}}{r}\right)^{12} - \left(\frac{\sigma_{wl}}{r}\right)^{6} f(x,y)\right] \quad (1)$$

where $\sigma_{wl}$ and $\varepsilon_{wl}$ are the LJ potential parameters between wall and liquid, r is the interparticle distance and $f(x,y)$ is given by Eq.(2).

$$f(x,y) = 1 + \frac{1}{3N}\sum_{K=1}^{3}\frac{1}{K}\left[\cos\left(\frac{4\pi Kx}{b}\right) + \cos\left(\frac{4\pi K(x+y\sqrt{3})}{2b}\right) + \cos\left(\frac{4\pi K(-x+y\sqrt{3})}{2b}\right)\right] \quad (2)$$

The parameter N controls the amplitude of the surface roughness. Smaller the value of N rougher is the surface. The parameter K represents the wave number of the surface characteristics whereas parameter b controls the periodicity of the function.

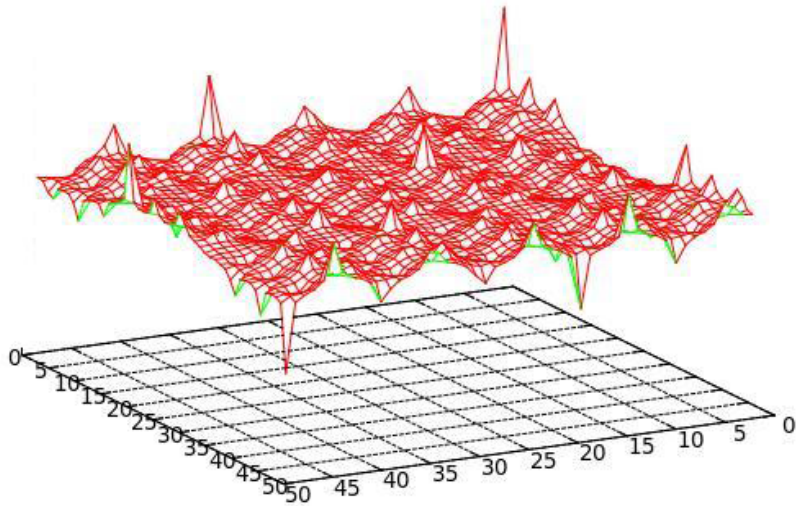

**Figure 2.** Simulated surface roughness ( $N = 2; \varepsilon_{ww} = 0.4; d = 10$ ): The grid represents the geometrical location of wall atoms; Surface represents the corrugations introduced by modification of LJ potential

Unlike the previous studies, choice of modified LJ potentials in this way not only allows a simultaneous control over the topographical and wettability characteristics of the confining boundaries through the pertinent interaction forces, but also implicates an explicit coupling between these two immensely consequential interfacial features. In these simulations, N is varied to get various degrees of surface roughness and fix $d$ such that an integral number of periods of roughness function are present in a simulation domain. This ensures continuity of roughness function across several simulation domains.

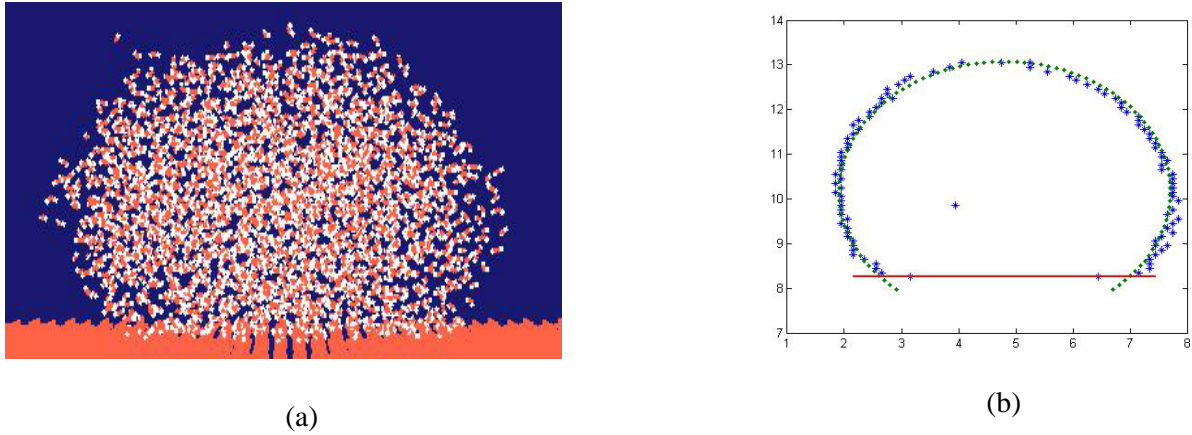

(a)                  (b)

**Figure 3.** Droplet surface as obtained by (a) Simulation ( $N = 2; \varepsilon_{ww} = 0.4$ ) and (b) Calculated iso-density curve and fitted circle ( $N = 2; \varepsilon_{ww} = 0.4$ )

The simulation system used for contact angle estimation is shown in Figure 1(b). In order to determine the contact angle, the surface of the droplet by joining all those points at which the density is half of the bulk density is defined and a circle is fitted through these points as shown in Figure 3(b). The angle between the surface and the tangent to the circle, at the intersection point of circle and surface, is the contact angle.

The simulations are performed using GROMACS MD package [24]. Production run for couette flow consisted of 2,000,000 steps with a time step of 0.001ps. Leap-frog algorithm was used for integration. Bond length and angle constraints were enforced with the SHAKE algorithm and a constant temperature of $298K$ was maintained with a Nose-Hoover thermostat. Periodic boundary conditions were used in XYZ directions.

**Results**

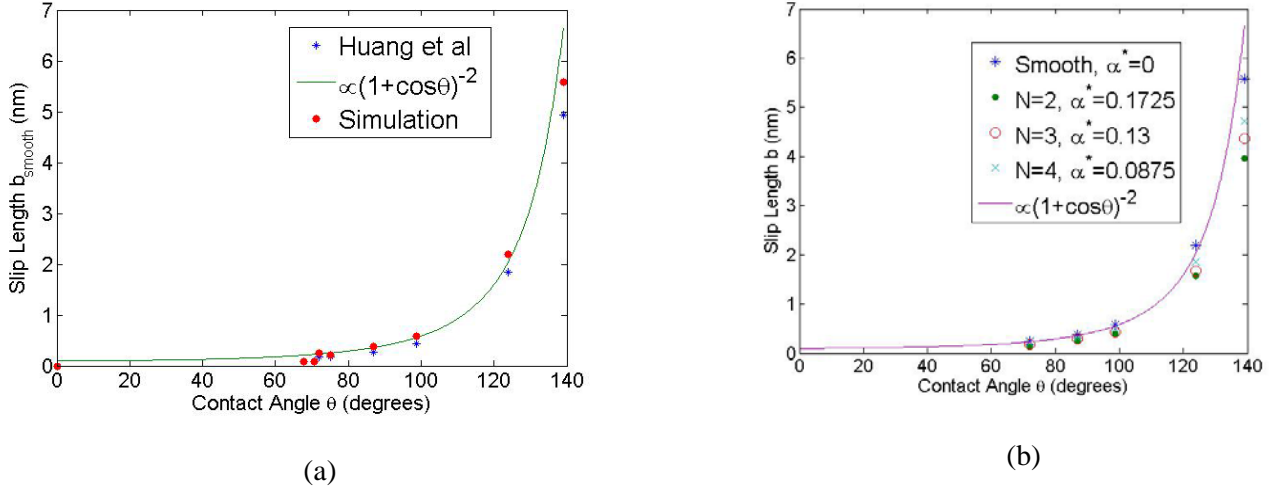

(a)    (b)

**Figure 4.** Slip length as a function of static contact angle for (a) Smooth LJ Surface and (b) Rough LJ Surfaces

Fig 4(a) shows the variation of slip length as a function of static contact angle. It is clear from the graph that the results of this work match closely with those obtained by Huang et al [21], confirming their proposition that slip length varies with the static contact angle as shown in Eq.(3)

$$b \propto (1+cos\theta)^{-2} \qquad (3)$$

where $b$ is the slip length and $\theta$ is the contact angle. Fig 4(b) shows the variation of slip length as a function of static contact angle for various degrees of surface roughness as obtained in the simulations. It is evident from the figure that for each value of roughness amplitude N, slip length shows the same variation with contact angle as suggested by Eq.(3). It is true that all the points, even for different surface roughnesses, do seem to lie on a universal curve, however, a general trend that slip length decreases with increasing surface roughness can clearly be observed, which warrants further investigation. For this purpose, the slip length $b$ is non-dimensionalized by the smooth surface slip length $b_{smooth}$, both obtained for the same value of wall-fluid interaction $(\varepsilon_{ww})$. The value of roughness function $f(x,y)$ not only represents the modification factor in LJ potential, but physically it is equivalent to the height (depth) of crests (troughs) on the surface. Mean value of $f(x,y)$ $\left(\overline{f(x,y)}\right)$ represents the height of the effective mean surface over the actual geometric surface. Therefore, one can define non-dimensional average surface roughness parameter $\alpha^*$ as shown in Eq.(4). $\alpha^*$ characterizes the roughness of a surface and varies only with N.

$$\alpha^* = \frac{\alpha}{\sigma_{wl}} = \frac{\sum |f(x,y) - \overline{f(x,y)}|}{N} \qquad (4)$$

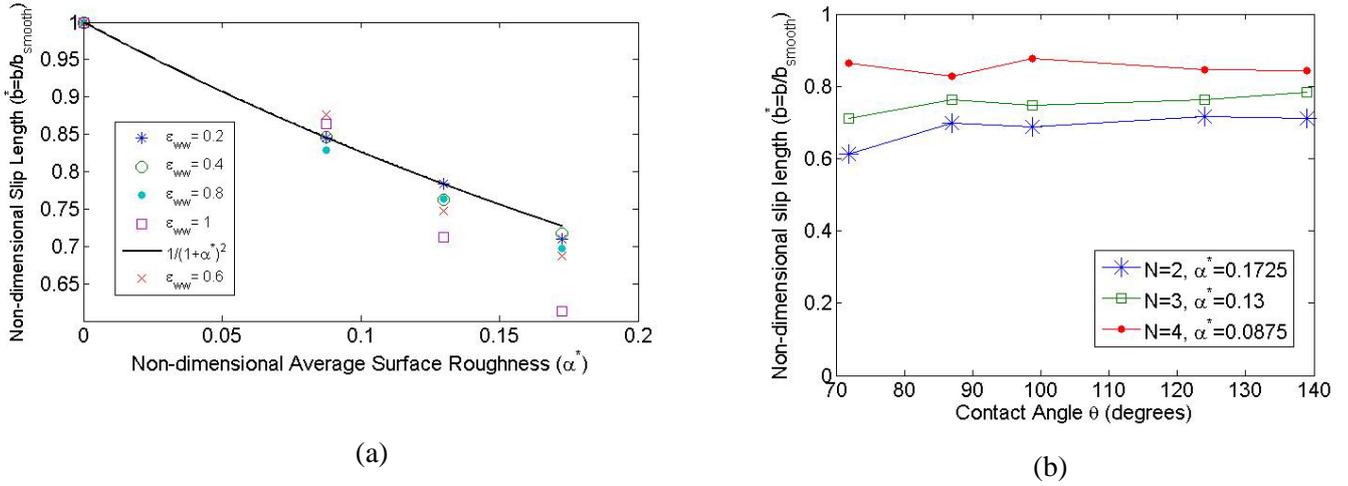

**Figure 5.** Non-dimensional slip length as a function of (a) Non-dimensional average surface roughness and (b) Contact angle

Non-dimensionalization of the slip length is done by dividing it by the hydraulic diameter of the channel. Fig. 5(a) shows this non-dimensionalized slip length as a function of non-dimensionalized average surface roughness for different values of wall-fluid interaction. Decreasing trend of slip length with increasing surface roughness can be clearly seen from the graph. Further, almost parallel curves indicate that qualitative dependence of slip length on non-dimensional average surface roughness is independent of wall-fluid interaction parameter $\epsilon_{ww}$.

Another length scale that can be used to non-dimensionalize the slip lengths obtained by varying the roughness for a particular value of $\epsilon_{ww}$ is the slip length obtained between fluid and smooth surface for that $\epsilon_{ww}$. Fig. 5(b) shows this non-dimensionalized slip length as a function of non-dimensionalized average surface roughness for different values of wall-fluid interaction ($\epsilon_{ww}$). Fig. 5(a) shows that irrespective of the value of $\epsilon_{ww}$, the non-dimensional slip length scales with non-dimensional average surface roughness as given by the Eq.(5). Deviation of curve for $\epsilon_{ww} = 1$ from Eq.(5), may be attributed to the hydrophilic nature of the wall seen at this value.

$$b^* = \frac{b}{b_{smooth}} = \frac{1}{(1+\alpha^*)^2} \qquad (5)$$

Huang et al [21], present a mathematical analysis through which they justify their universal scaling formula given in Eq.(3). Through this work, a correction to this scaling formula is proposed in order to calculate slip lengths for rough surfaces with more accuracy. The corrected formula is as shown in Eq.(6). Here k is the constant of proportionality which is same for both Eqns. (3) and (6).

$$b(\theta, \alpha^*) = \frac{k}{(1+cos\theta)^2 \times (1+\alpha^*)^2} \qquad (6)$$

One can verify the correctness of the formula presented in Eq.(5) through a simple mathematical analysis as shown below.

$$c_{wl}^6 = 4\epsilon_{wl}\sigma_{wl}^6 \Rightarrow c_{wl}^6 \propto \epsilon_{wl} \tag{7}$$

$c_{wl}^6$ being the attractive LJ interaction term and from Eq.(1),

$$c_{wl,rough}^6 = 4\epsilon_{wl}\sigma_{wl}^6 \times f(x,y) = c_{wl,smooth}^6 \times f(x,y) \tag{8}$$

$$b \propto \frac{1}{\epsilon_{wl}^2} \tag{9}$$

Therefore, combining Eqns. (7), (8) and (9)

$$\frac{b_{smooth}}{b_{rough}} = \left(\frac{\epsilon_{wl,rough}}{\epsilon_{wl,smooth}}\right)^2 = \left(\frac{c_{wl,rough}^6}{c_{wl,smooth}^6}\right)^2 = f(x,y)^2 \tag{10}$$

Since $f(x,y)$ varies over the surface, its average value is taken to represent the surface roughness. This average value, in non-dimensional form, is nothing but $\alpha^*$ as defined in Eq.(4). For a smooth surface $f(x,y)=1$ for all values of x and y, therefore $\alpha^*$ for a smooth surface is equal to 0. Since this value is calculated with respect to the mean value of $f(x,y)$, which is equal to 1 for all values of N (due to the symmetrical nature of $f(x,y)$ about the x-y plane), therefore 1 is added to $\alpha^*$ in order to obtain a true measure of $f(x,y)$. By substituting this measure of $f(x,y)$ in Eq.(10), the formula shown in Eq.(5) is obtained.

**Conclusions**

Through molecular dynamics simulations of a realistic water model, the slippage of water over smooth and rough surfaces is studied. By measuring slip lengths and contact angles under various conditions and studying their variations, a mathematical relationship between slip length, static contact angle, and average surface roughness is proposed. Finally, through a simple mathematical analysis, the theoretical basis of this two-way coupling of the chemical structure and physical roughness and the effect of the same on interfacial slip is established.